\documentclass[%
 reprint,
 amsmath,amssymb,
 aps,prl,superscriptaddress
]{revtex4-2}
\usepackage{xcolor}
\usepackage{graphicx}% Include figure files
\usepackage{dcolumn}% Align table columns on decimal point
\usepackage{bm}% bold math
\usepackage{amsmath}
\usepackage{physics}
\usepackage{siunitx}

\begin{document}

\preprint{APS/123-QED}

\title{Bragg Scattering from a   Random Potential}% Force line breaks with \\

\author{Donghwan Kim}
 
\affiliation{%
 Department of Chemistry and Chemical Biology, Harvard University, Cambridge, Massachusetts 02138, USA
}%

\author{Eric J. Heller}
\email{eheller@fas.harvard.edu}
\affiliation{%
 Department of Chemistry and Chemical Biology, Harvard University, Cambridge, Massachusetts 02138, USA
}%
\affiliation{%
 Department of Physics, Harvard University, Cambridge, Massachusetts 02138, USA
}%

\date{\today}% It is always \today, today,

\begin{abstract}

A potential for propagation of a wave in  two dimensions is constructed from a random superposition of plane waves around all propagation angles. Surprisingly, despite the lack of periodic structure, sharp Bragg diffraction of the wave is observed, analogous to a powder diffraction pattern. The scattering is partially resonant, so Fermi's golden rule does not apply. This phenomenon would be experimentally observable by sending an atomic beam into a chaotic cavity populated by a single mode laser.

%It is surprising as it is opposed to the wave scattering in liquid, e.g., x-ray scattering in water where broad peaks are observed in the scattering cross section. The broad peaks in the scattering cross section have corresponding broad peaks in the spatial autocorrelation function of atoms. Thus, the former is used to infer the latter and vice versa. However, in the wave scattering in the weak random potential, spatial autocorrelation is no longer useful to predict the scattering cross section. Instead, looking at the Fourier components of the potential is crucial in understanding the scattering phenomena including cross section. In addition, the scattering rate cannot be explained by Fermi's golden rule. Fermi's golden rule is designed for explaining scattering from a single state to a continuum, but the scattering in the weak random potential is from a single state to another single state, so Fermi's golden rule does not work.

%This could be verified experimentally by using Bose-Einstein condensate in a randomly scattered single mode laser field (compact name for it?).

\end{abstract}

\maketitle

%\tableofcontents

% \section{\label{sec:level1}Introduction}
% \begin{figure}
%     \centering
%     \includegraphics[width=3in]{autocorr.JPG}
%     \includegraphics[width=3in]{strucfun.JPG}
%     \caption{Look at $\SI{200}{\celsius}$ data. The broad peak near $\SI{2.8}{\angstrom}$ in the autocorrelation function (top figure) gives a broad peak near $\SI{2.5}{\angstrom^{-1}}$ in the structure function (bottom figure). The figures were copied from \cite{doi:10.1063/1.1676403}.}
%     \label{fig:XrayWater}
% \end{figure}
Monochromatic light or matter waves entering a perfectly periodic medium show sharp Bragg scattering  into specific angles. However, randomly   disturbing the perfect lattice positions results in diffuse scattering between the Bragg peaks. As the dispersion increases,  the diffuse scattering eventually dominates and finally the Bragg peaks vanish.
The diffuse scattering    is  structured, revealing   correlations in the medium. 
For example,  for x-ray scattering in water \cite{DF9674300097,doi:10.1063/1.1676403} and the scattering of visible light in disordered packing of monodisperse polystyrene beads \cite{PhysRevA.102.033529,PhysRevE.101.012614}, the pair correlation function has a broad peak with a characteristic length scale, which in turn generates a broad peak in the structure function.  %as shown in FIG. \ref{fig:XrayWater}.

In the studies of disordered media, the Bragg peaks are associated with periodic structures  \cite{PRLSegev,pnasF}.
It is not expected, however, that a random medium, with no perfect order on any scale, can generate  sharp scattering angles, yet we report such a case here. 
For the potential we choose, the spatial autocorrelation function  has broad peaks as the atom pair correlation function in water, but the scattering angle nonetheless is very sharp. This is startling;  the scattering in the random potential defined below is   like  Bragg scattering in a periodic potential,   rather than the scattering in a correlated liquid. The closest analog---though not a perfect one---is powder diffraction with many randomly oriented crystallites packed closely.  The potential defined below  has no such ``crystallites,'' yet it has Bragg peaks. 
We explain this surprise by   calculating scattering matrix elements, or, equivalently, by  examining the Fourier components of the potential.  However, the time evolution of the scattering is not compatible with  Fermi's golden rule, as discussed below.  
% Given an incident wave packet of well-defined momentum, the population growth of the Bragg-like  scattered wave is oscillatory, like a resonant double well. The Bragg-like scattering develops quickly, as befits a resonant process.  Starting with a wave packet having a well-defined direction of propagation, sporadic ``Bragg pulses" emerge from it (see e.g. FIG. \ref{BraggBerry} and \ref{hbars}).

% \section{Theory and Simulation}
% \subsection{Definition of the random potential}
We consider the following form of random potential
\begin{align}
    V(\Vec{r};\{\phi_j\})=\frac{A}{\sqrt{N}}\sum_{j=1}^N\cos(\Vec{q}_j\cdot\Vec{r}+\phi_j)
    \label{eq:randompot}
\end{align}
% where $A$ is a constant having the dimension of energy, $N$ is the number of modes, $\Vec{q}_j=q(\theta_j)(\hat{x}\cos\theta_j+\hat{y}\sin\theta_j)$, and $\theta_j=2\pi(j-1)/N$. This is a superposition of $N$   equal amplitude plane waves each propagating in a direction with an angle $\theta_j$, a wave number $q(\theta_j)$ depending on the angle, and a random phase shift $\phi_j$. 
% A special case of this potential has all the wavelengths of the modes the same $q(\theta_j)=q$; this we call  a ``Berry potential,''  a function introduced in connection with wave chaos \cite{Berry_1977, longuet}.
where $A$ is a constant having the dimension of energy, $N$ is the number of modes, $\Vec{q}_j=\abs{\Vec{q}_j}(\hat{x}\cos\theta_j+\hat{y}\sin\theta_j)$ are wave vectors, and $\theta_j=2\pi(j-1)/N$ are angles equally spaced over $2\pi$. This is a superposition of $N$ plane waves of an equal amplitude $A$ each propagates in different directions with a wave number $\abs{\Vec{q}_j}$, an angle $\theta_j$ and a random phase shift $\phi_j$. 
% The normalization factor $1/\sqrt{N}$ is introduced, making the average fluctuation $V_{\textrm{rms}}=\sqrt{\ev{(V(\Vec{r}))^2}}$   independent of the number of modes $N$. 
It is not important that the angle be equally spaced, or the amplitudes be the same, as long as they are random.

For simplicity, we consider a random potential constructed by equal wave number $\abs{\Vec{q}_j}=q$ (see Supplemental Material for more general potentials).
We call this a ``Berry potential,''  a function introduced in connection with wave chaos \cite{Berry_1977, longuet}.
This random potential is experimentally realizable in a laser cavity with a single mode laser (need not be single) with rough or ballistically chaotic walls, so that the wave inside is a random superposition of waves traveling in all directions.

% This potential  could be produced for example in atom optics by a single mode laser (with constant $\abs{\Vec{q}_j}$ ) in a cavity with walls that scramble  scattering angles.  This form is also one component of the thermal deformation potential, which has a   Bose-Einstein distribution with a Debye cut-off in  $\abs{\Vec{q}_j}$.

% It is not important to take the   propagation directions  to be distributed evenly around $2 \pi$. The directions could be chosen randomly, as could the amplitudes $A$, becoming $A_j,$   drawn from a distribution.  These variations lead to statistically indistinguishable functions in the large $N$ limit, by the central limit theorem.  

In the limit of many modes $N\to\infty$, the spatial autocorrelation of the potential is (see Supplemental Material)
\begin{align*}
    C(\delta\Vec{r})
    =
    \ev{V(\Vec{r})V(\Vec{r}+\delta\Vec{r})}
    =
    \frac{A ^2}{2} J_0(q\delta r)
\end{align*}
%The fluctuation of the potential $V_{\textrm{rms}}=\sqrt{C(0)}=A/\sqrt{2}$ is indeed independent of the number of modes $N$.
% \begin{align*}
%     C(\delta\Vec{r})
%     &=
%     \frac{A ^2}{N\Delta\theta}\sum_{j=1}^N\Delta\theta\cos(\Vec{q}_j\cdot\delta\Vec{r})/2
%     \\
%     &\approx
%     \frac{A ^2}{2\pi}\int_0^{2\pi}\dd\theta\cos(q\delta r\cos\theta)/2
%     \\
%     &=
%     \frac{A ^2}{2} J_0(q\delta r)
% \end{align*}
where $J_0$ is the zeroth-order Bessel function of the first kind. 
% The autocorrelation function of the Berry potential is shown in FIG. \ref{BerryACPot}. %Note the peaks give characteristic length scales shown in the potential plot in FIG. \ref{BerryPot}.
From the autocorrelation, we can obtain the root-mean-square of the potential
\begin{align*}
    V_{\textrm{rms}}=\sqrt{\ev{(V(\Vec{r}))^2}}=\sqrt{C(0)}=A/\sqrt{2}.
\end{align*}

\begin{figure}
    \centering
    \includegraphics[width=3.4in]{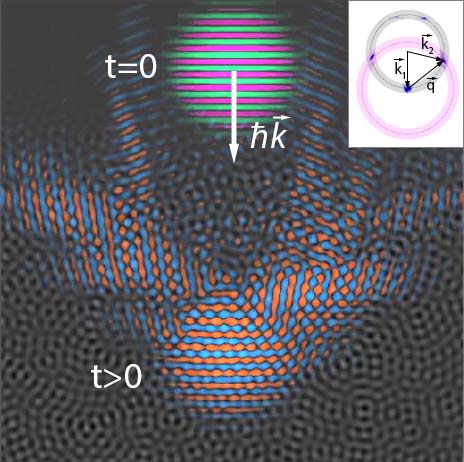}
    \caption{Bragg scattering of a wave packet in a random (Berry) potential. The potential is drawn in gray scale on the background. An initial wave packet on the top launched downward into the potential with an average momentum $\hbar\Vec{k}$ is depicted in green and magenta scale. The wave function at a later time is shown in red and blue scale.
    The inset shows the probability density of the wave at a later time in the reciprocal space in dark blue scale.}
    \label{BraggBerry}
\end{figure}

\begin{figure*}
    \centering
    \includegraphics[width=.32\textwidth]{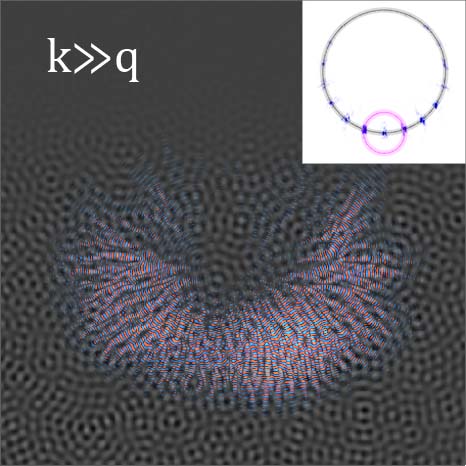}
    \includegraphics[width=.32\textwidth]{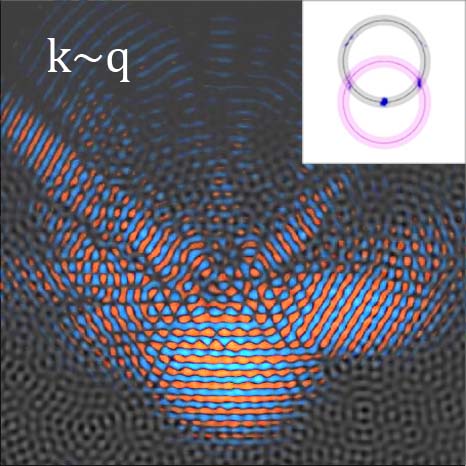}
    \includegraphics[width=.32\textwidth]{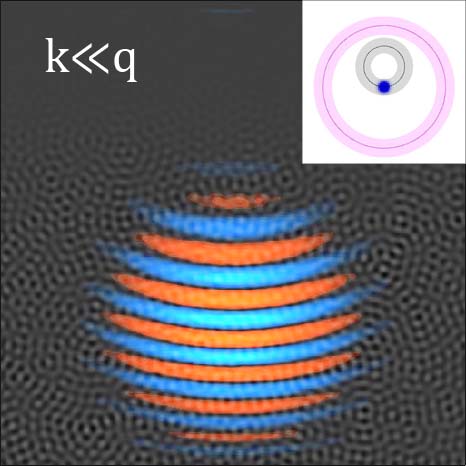}
    \caption{Dependence of the wave evolution  on  $\hbar$. The real part of the wave function is plotted in red and blue scale and an inset in  each panel shows the probability density distribution in the reciprocal space. For small $\hbar$, i.e., $k\gg q$ (small wavelength and classical limit), the wave dynamics is particlelike, diffusive, showing branched flow \cite{branchedFlow}. This is consistent with a very small Bragg angle which leads to repeated almost-forward scattering.
    For intermediate $\hbar$, i.e., $k\sim q$ the Bragg angle is not small, giving less classical-like  diffractive behavior.
    For large $\hbar$, i.e., $k\ll q$ the wave is in a transparency regime:  the wavelength is   large enough that the small scale fluctuations of the potential are   averaged to zero. Or equivalently, in reciprocal space, there are no intersections between the energy contour and the nonzero   Fourier components of the potential. Therefore, there  is effectively no scattering, and the potential is transparent.}
    \label{hbars}
\end{figure*}

\begin{figure}
    \centering
    \includegraphics[width=3.3in]{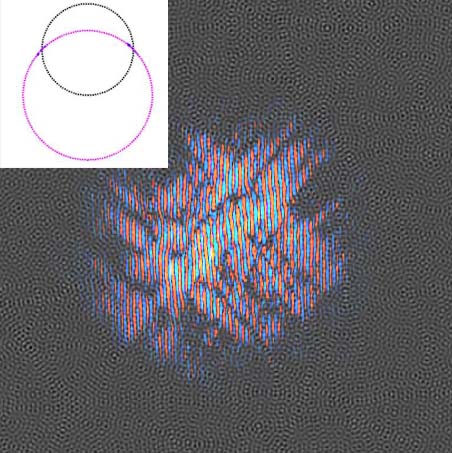}
    \caption{The real part of first-order wave function $\psi^{(1)}(\Vec{r},t)$ from a wave packet initially launched downward, showing the interference of the instantaneous scattered amplitudes from different times.}
    \label{1stPT}
\end{figure}

% \subsection{Simulation method}
% Now we propagate wave packets of fairly well defined momentum $\hbar\Vec{k}$ in the potential $ V(\Vec{r})$. 
% We use a  second order split operator method \cite{SplitOperator} where the time evolution operator for the   Hamiltonian $H=T+V$ where $T$ and $V$ are kinetic and potential energy operators, respectively, is approximated as
% \begin{align*}
%     e^{-iH\Delta t/\hbar}
%     =
%     e^{-iV\Delta t/2\hbar}
%     e^{-iT\Delta t/\hbar}
%     e^{-iV\Delta t/2\hbar}
%     +
%     \mathcal{O}((\Delta t)^3)
% \end{align*}
% The propagation of a wave function in real space $\psi(\Vec{r},t_0)$ at time $t_0$ is implemented by  Fourier transform alternating between position and momentum space. Since the kinetic and potential energy operators depend only on momentum and position operators, respectively, i.e., $T=T(\hat{\Vec{p}})$ and $V=V(\hat{\Vec{r}})$, it is easier to Fourier and inverse Fourier transform the wave function in the following way during the computation
% \begin{align*}
%     \psi(\Vec{r},t_0+\Delta t)
%     &=
%     e^{-iV(\hat{\Vec{r}})\Delta t/2\hbar}
%     \mathcal{F}^{-1}\{
%     e^{-iT(\hat{\Vec{p}})\Delta t/\hbar}
%     \\
%     &\qquad
%     \mathcal{F}[
%     e^{-iV(\hat{\Vec{r}})\Delta t/2\hbar}
%     \psi(\Vec{r},t_0)
%     ]\}
%     +
%     \mathcal{O}((\Delta t)^3)
% \end{align*}
% where $\mathcal{F}$ is the Fourier transform operator from position to reciprocal space.

% \subsection{Bragg scattering from the potential and an analogy with powder x-ray scattering}
We study the dynamics of wave packets with an initial average momentum $\hbar k$ in the Berry potential $V(\Vec{r})$ employing the second order split operator method~\cite{SplitOperator}.
To have a ``weak'' disorder strength, set the constant $A$ such that the fluctuation of the potential $V_{\textrm{rms}}$ is far smaller than the average kinetic energy of the wave $\ev{T}=\frac{\hbar^2k^2}{2m}$, i.e., $V_{\textrm{rms}}\ll \ev{T}$. 
Consideration of strong disorder strength $V_{\textrm{rms}}\gtrsim \ev{T}$ is for future study \cite{locnote}.%,*donghwan}.

% In powder diffraction from three dimensions  crystallites, the  Bragg scattering forms cones of discrete angular deviation from incident direction.  For a two dimensional ``powder'', the allowed directions for in-plane propagation are reduced to discrete in-plane angles, as we see in our studies.

Fig. \ref{BraggBerry} shows the propagation of a wave packet in the Berry potential. 
The potential is drawn in gray scale on the background. There is a flat zero potential on the top and the Berry potential is smoothly turned on toward the bottom.  An initial wave packet on the top launched downward into the potential with an average momentum $\hbar\Vec{k}$ is depicted in green and magenta scale. The wave function at a later time is shown in red and blue scale. The wave is scattered by the random potential, initially only to  the Bragg angle, but this is soon scattered again with the same Bragg angle relative to the motion, and so on.
This  scattering should not be confused with the higher order Bragg scattering that is absent if the Berry potential is composed only of sinusoids. The higher order scattering would be allowed if instead triangular waves were used, for example.

% The inset in Fig. \ref{BraggBerry} shows the probability density of the wave at a later time in the reciprocal space in dark blue scale. The initial $\Vec{k_1}$, final $\Vec{k_2}$, scattering $\Vec{q}$ wave vectors are shown. For a given initial wave vector $\Vec{k_1}$, the contour of all equal energies can be drawn as a black circle. Taking $\Vec{k_1}$ as an origin, the nonzero Fourier components of the potential are drawn as a magenta band.
% (The initial wave packet has a small momentum dispersion, so the circles become bands). The Bragg scattered amplitudes appear over small ranges,  determined by the intersections of the black and magenta bands.%$\Vec{k_1}$ can be chosen to be any wave vector near the initial average value $\Vec{k}$. The shaded gray and magenta area are the area covered by the collection of the black and magenta circles, respectively, considering all of the initial wave vector $\Vec{k_1}$'s near the average value $\Vec{k}$. 
% In the reciprocal space, it is  seen that the scattered waves are populated only at Bragg angles.

The inset in Fig. \ref{BraggBerry} shows the probability density of the wave at a later time in the reciprocal space in blue scale. The initial $\Vec{k_1}$, final $\Vec{k_2}$, and scattering $\Vec{q}$ wave vectors are shown as in Ewald sphere construction \cite{ewald1969introduction,ashcroft1976solid}. For a given initial wave vector $\Vec{k_1}$, the contour of equal energy can be drawn as a black circle. In addition, taking $\Vec{k_1}$ as an origin, the nonzero Fourier components of the potential can be drawn as a magenta circle.
Then, the Bragg scattered states appear at the intersections of the black and magenta circles \cite{fignote}.
It is seen that the scattered waves are populated only at a Bragg angle.
In three dimensions, the construction involves the intersection of spheres and thus the outgoing wave vectors $\Vec{k_2}$ will lie on a ring corresponding to a Bragg angle.

The Bragg scattering of the wave in the Berry potential can roughly be interpreted as a superposition of the scattered waves by each constituent sinusoidal potential aligned in different directions. %Most of those directions do not scatter for a given incident nearly-plane wave.
The situation is analogous to  powder x-ray diffraction in which crystallites are aligned in all possible orientations, leading to the incoherent superposition of the outgoing waves from scattering by the crystallites. The conventional picture is that the scattering by some of the crystallites oriented properly with respect to incident x-ray beam leads to Bragg scattering,  although this view has been challenged \cite{fewster_2014} in ways that are relevant to our present observations.  Again there are no crystallites here, but the Berry potential bears some relation to the impossible limit of overlapping and blending them. In three dimensions, the scattered waves from the Berry potential will form a ring as is the case in the powder diffraction.

Employing the analogy above, by treating the wavelength $2\pi/q$ of the single sinusoid in the Berry potential as a ``lattice spacing,'' one can write down the Bragg condition $n(2\pi/k)=2(2\pi/q)\sin\theta$, which correctly explains our simulation result. Note only the first order ($n=1$) Bragg angle $\theta_B$ satisfying $(2\pi/k)=2(2\pi/q)\sin\theta_B$ is observed. Higher order ($n=2,3,\dots$) Bragg angles will be observed if triangular, instead of sinusoidal, waves are used.
% Let us denote the first order ($n=1$) Bragg angle as $\theta_B=\sin^{-1}(q/2k)$.
Nevertheless, this analogy is not perfect since, in the random Berry potential, the superposition of the scattered waves by each sinusoidal component is coherent, rather than incoherent which was the case in the powder diffraction as the phases of scattered waves from one crystallite to another do not match. The Berry potential coherence effect is manifested in the rapid  growth of the scattered wave population as discussed below.

% \subsection{Analysis in reciprocal space.}
Consider scattering matrix element $\mel{\Vec{k_2}}{V}{\Vec{k_1}}\sim V_{\Vec{k_2}-\Vec{k_1}}$ where $\ket{\Vec{k_1}}$ and $\ket{\Vec{k_2}}$ are plane wave states and $V_{\Vec{k_2}-\Vec{k_1}}$ is the Fourier component of the potential. 
%Also, in the first order time-dependent perturbation theory, as one can see from Eq.\ref{TDPT}, the scattered waves as energy conserving transition is favored more as time goes by, elastic scattering dominates.
For elastic scattering ($\abs{\Vec{k_1}}=\abs{\Vec{k_2}}=k)$, the scattering wave vector $\Vec{q}=\Vec{k_2}-\Vec{k_1}$ and scattering angle $2\theta$ satisfy $q=2k\sin\theta$, which coincide with the Bragg condition $n(2\pi/k)=2(2\pi/q)\sin\theta$ for $n=1$.
Thus, allowed elastic scattering  can actually be interpreted as  Bragg scattering by a plane wave ``lattice'' of one sinusoidal component of the potential properly aligned with respect to the incident beam direction.
% Thus,  the Bragg scattering  can be interpreted as  diffraction  by a plane wave ``lattice'' of one sinusoidal component of the potential properly aligned (and amplified in magnitude) with respect to the incident beam direction. 
The random Berry ``medium'' has no special directions of travel; all are equivalent and subject to Bragg diffraction at {\it relative} angle $\theta$.

% Using the scattering matrix element analysis, one can draw nonzero Fourier components of the potential on the reciprocal space probability distribution plot as shown in FIG. \ref{BraggBerry}. The simulation result shows the reciprocal space picture explains the scattering phenomenon correctly.
% The momentum dispersion of the initial wave packet makes a small range of Bragg angles populated as explained in the caption of FIG. \ref{BraggBerry}. % Indeed every momentum component of the wave follows the Bragg condition, with larger momentum (slightly outside the reference momentum circle) having smaller Bragg angles.

% \subsection{Length scales of the wave and potential}

The scattering behavior of the wave varies considerably  depending on its wavelength $2\pi/k$ compared to the wavelength of the sinusoids $2\pi/q$ used in the Berry potential. 
To understand the effect of length scales, we compare wave scattering for different values of reduced Planck's constant $\hbar$, keeping the average momentum of the wave $\hbar k$ fixed. 
% Thus if the value of $\hbar$ increases by a factor of two, the wave number of the wave $k$ decreases by a factor of two,  and the wavelength increases by two, keeping the average momentum fixed.
By keeping the momentum the same, the wave propagation speed and kinetic energy are kept the same for different $\hbar$'s, so the only physical difference comes from different wavelengths.

Depending on the value of $\hbar$ (so the wavelength), the wave packet scattering  exhibits qualitatively different behaviors as shown in Fig. \ref{hbars}. For small $\hbar$, i.e., $k\gg q$ (small wavelength and classical limit), the wave dynamics is particlelike, diffusive, showing branched flow \cite{branchedFlow}. This is consistent with a very small Bragg angle which leads to almost-forward scattering.
For intermediate $\hbar$, i.e., $k\sim q$ the Bragg angle cannot be treated to be small which results in less classical looking and diffractive behavior.
For large $\hbar$, i.e., $k\ll q$ the wave is in a transparency regime;  the wavelength is   large enough that the smaller scale fluctuations of the potential are   averaged to zero. Equivalently, in reciprocal space, there are no intersections between the energy contour and the nonzero   Fourier components of the potential. Therefore, there  is effectively no scattering, and the potential is transparent.

The first-order time-dependent perturbation theory gives the first-order correction to the wave function \cite{Heller:way}
\begin{align}
    \psi^{(1)}(\Vec{r},t)=\int_0^t\phi(\Vec{r},t')\dd t'
    \label{TDPT}
\end{align}
where the instantaneous scattered amplitudes $\phi(\Vec{r},t')=\frac{1}{i\hbar}e^{-iT(t-t')/\hbar}V(\Vec{r})e^{-iTt'/\hbar}\psi(\Vec{r},0)$ at different times $t'$ interfere to form  $\psi^{(1)}(\Vec{r},t)$. Fig. \ref{1stPT} shows the first-order wave function $\psi^{(1)}(\Vec{r},t)$ from the wave packet initially launched downward. One can see the interference pattern formed by the superposition of $\phi(\Vec{r},t')$ at different times $t'$.
Also, the first-order wave function $\psi^{(1)}(\Vec{r},t)$ depends on the size and shape of the initial wave function $\psi(\Vec{r},0)$ and the region of the potential the wave is spatially lying on. The asymmetry of the scattered waves shown in Figs. \ref{BraggBerry}, \ref{hbars}, and \ref{1stPT} is due to the asymmetry of the region of the potential right underneath the wave. This is not captured by the usual plane wave perturbation theory which predicts the symmetry of the scattered waves.

\begin{figure}
    \centering
    \includegraphics[width=3.5in]{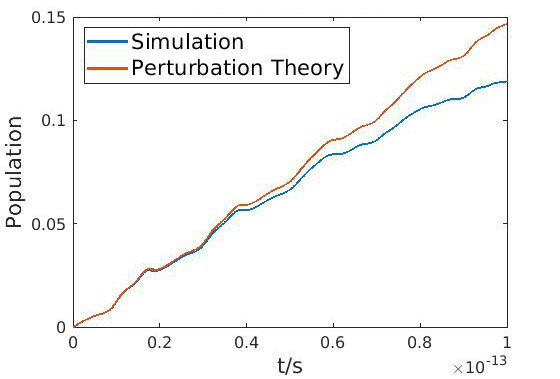}
    \caption{Population of the scattered waves as a function of time. Blue and red curves correspond to simulation and 1st order time-dependent perturbation theory results, respectively. The population growth is not linear, i.e., Fermi's golden rule is not valid. Also, the population growth stops occasionally, showing that the pulses (scattered waves) come out sporadically as shown in Figs. \ref{BraggBerry} and \ref{hbars}.
    % Top panel is the result from a very weak potential. The simulation and the perturbation theory results coincide. The bottom panel is the result from a stronger, but still weak potential.
    The two results coincide at  early times when single scattering dominates. The discrepancy later is due to multiple scattering, absent in the 1st order time-dependent perturbation theory.}
    \label{PopulationGrowth}
\end{figure}

One can calculate the population $\braket{\psi^{(1)}(t)}$ of the scattered waves from the first-order perturbation theory.
The populations near the Bragg angle $\theta_B$ (so the scattering angle $2\theta_B$) were calculated in the simulation and compared with the perturbation theory as shown in Fig. \ref{PopulationGrowth}.
The nonlinear population growth indicates the breakdown of Fermi's golden rule. 
This behavior combines  aspects of both resonant and nonresonant decay, which depends on the specifics of the interference of the scattered amplitudes $\phi(\Vec{r},t')$ in space and time~\cite{Heller:way}.
Furthermore, the sometimes strong and irregular population growth oscillations are captured in the first-order time-dependent perturbation theory,  explaining the sporadic pulses (scattered wave) coming out as shown in Figs. \ref{BraggBerry} and \ref{hbars}.
Again, the plane wave perturbation theory does not correctly predict the population growth.

It is worth emphasizing the explanations given are not restricted to the two dimensional Berry potential, but are valid for more complicated forms of random potentials and in three dimensions as well. We checked the validity for more general two dimensional random potentials where the wave numbers $\abs{\Vec{q}_j}$ of the modes are different and even forming a ``band'' in reciprocal space (see Supplemental Material).
Also, as the explanations do not employ any specific property of two-dimensionality, they are expected to be valid in three dimensions as well.

% This is because Fermi's golden rule assumes (1) the scattering to be from a single state to a continuum, and (2) the superposition of the scattered waves to be incoherent. Both are not true in this case; the coupling is from a single state {\it coherently} to a set of degenerate states. The coupling matrix elements vary considerably.  

% The population growth and scattering behavior depend   on the size, shape and speed of the initial wave. 

%In addition, this effect arises since the nonzero components of the potential in Eq. \eqref{eq:randompot} populates a one dimensional manifold in the Fourier space plane.
% If it starts filling areas, like the deformation potential mentioned below, the effect will go away at some rate.
%If it starts filling areas, the effect will go away at some rate.

\begin{figure}
    \centering
    \includegraphics[width=3in]{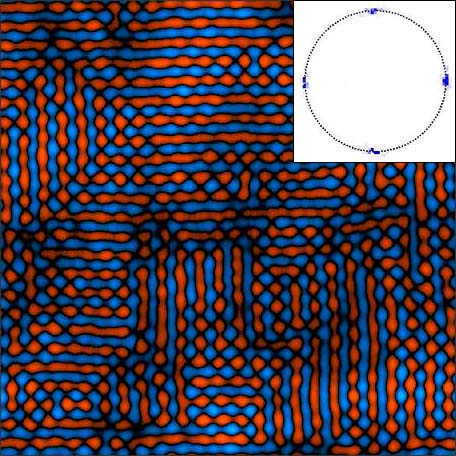}
    \caption{Bragg scattering in a Berry potential: a special case of the 90$^\circ$ scattering angle. The real part of the wave function was plotted in red and blue scale.  The inset shows the momentum space probability distribution of the wave. This is a snapshot, after a long time propagation of a downward launched wave packet. It is not diffusing in angle beyond the 90$^\circ$ turns, in spite of the random nature of the potential. A periodic boundary condition is used.}
    \label{90Bragg}
\end{figure}

% \subsection{Localization of the waves in momentum space}

Interestingly, there exists a momentum localization for special Bragg angles as shown in Fig.~\ref{90Bragg}. A snapshot of the wave is shown  after long propagation of a plane wave launched downward, in a 90$^\circ$ scattering (45$^\circ$ Bragg angle)  situation.  Of course, the scattered wings again rescatter at 90$^\circ$, and so on.  There may result a permanent  localization only to vertical and horizontal motion.

If the diffraction angle is $2\pi/N$ where $N$ is a positive integer, the scattered wave comes back to the original incident angle after $N$ scatterings, so the momentum distribution does not fill in the whole range of $2 \pi$.  The special angles show localization of the wave in momentum space, assuming the initial wave packet is narrow enough in momentum space. In three dimensions, the situation is a little bit different from two dimensions, and the momentum localization will be possible only for scattering angle $\pi$~\cite{3DLocalization}.

We would like to emphasize that the presented results apply to any linear wave transport: not only to matter waves, but also to acoustic and electromagnetic waves.

The Berry potential may be used as a diffraction grating in the diffractive regime. If the incident beam is white, after passing the Berry potential, it will be broken into its constituent colors since different colors have different Bragg angles. The difference between the usual diffraction gratings and the Berry potential is that the diffraction pattern from the latter will be independent of the incident beam directions since the Berry potential is isotropically random.

In conclusion, although a random (Berry) potential lacks periodicity, sharp Bragg diffraction of the wave is observed in the potential, analogous to a powder diffraction pattern. Fermi's golden rule breaks down since the scattering is partially resonant. This phenomenon would be experimentally realizable by sending an atomic beam into a chaotic cavity populated by a single mode laser. 

MATLAB codes implementing the split operator method can be found in Ref. \cite{MATLABcode}.

\begin{acknowledgments}
The authors thank David Gu{\'e}ry-Odelin, Alvar Daza, Frans A. Spaepen, Joonas Keski-Rahkonen, Paul F. Fewster and Joon-Suh Park for discussion and suggestion. We thank the NSF Center
for Integrated Quantum Materials (CIQM) Grant No.
DMR-1231319.
\end{acknowledgments}

\bibliography{apssamp}% Produces the bibliography via BibTeX.

\begin{thebibliography}{18}
\bibitem [{fig()}]{fignote}%
  \BibitemOpen
  \href@noop {} {}\bibinfo {note} {As the initial wave function has the
  momentum spread, $\Vec{k_1}$ can be chosen to be any wave vector near the
  initial average value $\Vec{k}$. The shaded gray and magenta area are the
  area covered by the collection of the black and magenta circles,
  respectively, considering all of the initial wave vector $\Vec{k_1}$'s near
  the average value $\Vec{k}$.}\BibitemShut {Stop}%
\bibitem [{3DL()}]{3DLocalization}%
  \BibitemOpen
  \href@noop {} {}\bibinfo {note} {In three dimensions, the scattering angle of
  $\pi/2$ will first fill the great circle of the equal energy sphere in
  reciprocal space. Then, the scattered waves on the great circle will be
  scattered again to fill the whole sphere, losing the localization feature.
  This is why only scattering angle of $\pi$ will lead to localization in the
  reciprocal space in three dimensions.}\BibitemShut {Stop}%
\end{thebibliography}

\end{document}

\preprint{APS/123-QED}

\title{Bragg scattering from a   random potential: Supplemental material}% Force line breaks with \\

\author{Donghwan Kim}
 
\affiliation{%
 Department of Chemistry and Chemical Biology, Harvard University, Cambridge, Massachusetts 02138, USA
}%

\author{Eric J. Heller}
\email{eheller@fas.harvard.edu}
\affiliation{%
 Department of Physics, Harvard University, Cambridge, Massachusetts 02138, USA
}%
\affiliation{%
 Department of Chemistry and Chemical Biology, Harvard University, Cambridge, Massachusetts 02138, USA
}%

\date{\today}% It is always \today, today,

\maketitle
\tableofcontents
\section{Spatial autocorrelation function of the potential}
For the random potential in the main manuscript
\begin{align}
    V(\Vec{r})=\frac{A}{\sqrt{N}}\sum_{j=1}^N\cos(\Vec{q}_j\cdot\Vec{r}+\phi_j),
    \label{eq:randompot}
\end{align}
the spatial autocorrelation function of the potential is
\begin{align}
    C(\delta\Vec{r})
    &=
    \ev{V(\Vec{r})V(\Vec{r}+\delta\Vec{r})}
    \\
    &=
    \frac{A ^2}{N}\sum_{j,j'=1}^N\ev{\cos(\Vec{q}_j\cdot\Vec{r}+\phi_j)\cos(\Vec{q}_{j'}\cdot(\Vec{r}+\delta\Vec{r})+\phi_{j'})}
    \label{eq:AC1}
    \\
    &=
    \frac{A ^2}{N}\sum_{j,j'=1}^N
    [\ev{\cos((\Vec{q}_j+\Vec{q}_{j'})\cdot\Vec{r}+\phi_j+\phi_{j'}+\Vec{q}_{j'}\cdot\delta\Vec{r})}
    +\ev{\cos((\Vec{q}_{j'}-\Vec{q}_j)\cdot\Vec{r}+\Vec{q}_{j'}\cdot\delta\Vec{r}+\phi_{j'}-\phi_j)}]/2
    \label{eq:trig}
\end{align}
where we used trigonometric identity
$\cos{\alpha}\cos{\beta}=[\cos(\alpha+\beta)+\cos(\alpha-\beta)]/2$.
The average $\ev{\cdot}$ can be taken over either position $\Vec{r}$
\begin{align*}
    \ev{\cdot}
    =
    \mathcal{A}^{-1}\int_{\mathcal{A}}\dd\Vec{r}\ \cdot
\end{align*}
where $\mathcal{A}$ is the area of the chosen region,  or random phases $\{\phi_j\}_j$
\begin{align*}
    \ev{\cdot}
    =
    (2\pi)^{-N}\int_{0}^{2\pi}\dd\phi_1\int_{0}^{2\pi}\dd\phi_2\cdots\int_{0}^{2\pi}\dd\phi_N\ \cdot,
\end{align*}
which is basically disorder average or average over potential realizations.
They all give the identical results as both the position $\Vec{r}$ and random phases $\{\phi_j\}_j$ appear in the argument of the same cosine in Eq. \eqref{eq:randompot}.
The first term in the bracket of Eq. \eqref{eq:trig} vanishes as one can check from position average
\begin{align*}
    \mathcal{A}^{-1}\int_{\mathcal{A}}\dd\Vec{r}\cos((\Vec{q}_j+\Vec{q}_{j'})\cdot\Vec{r}+\phi_j+\phi_{j'}+\Vec{q}_{j'}\cdot\delta\Vec{r})
    =
    0
\end{align*}
given the area $\mathcal{A}$ is large enough to cancel out the oscillations of the cosine, or from disorder average
\begin{align*}
    (2\pi)^{-N}\int_{0}^{2\pi}\dd\phi_1\int_{0}^{2\pi}\dd\phi_2\cdots\int_{0}^{2\pi}\dd\phi_N\cos((\Vec{q}_j+\Vec{q}_{j'})\cdot\Vec{r}+\phi_j+\phi_{j'}+\Vec{q}_{j'}\cdot\delta\Vec{r})
    =
    0.
\end{align*}
Likewise, the second term in the bracket of Eq. \eqref{eq:trig} vanishes if $j\neq j'$ as one can check from position or disorder average
\begin{align*}
    \mathcal{A}^{-1}\int_{\mathcal{A}}\dd\Vec{r}\cos((\Vec{q}_{j'}-\Vec{q}_j)\cdot\Vec{r}+\Vec{q}_{j'}\cdot\delta\Vec{r}+\phi_{j'}-\phi_j)
    =
    \delta_{j,j'}\cos(\Vec{q}_{j}\cdot\delta\Vec{r})
    \\
    (2\pi)^{-N}\int_{0}^{2\pi}\dd\phi_1\int_{0}^{2\pi}\dd\phi_2\cdots\int_{0}^{2\pi}\dd\phi_N\cos((\Vec{q}_{j'}-\Vec{q}_j)\cdot\Vec{r}+\Vec{q}_{j'}\cdot\delta\Vec{r}+\phi_{j'}-\phi_j)
    =
    \delta_{j,j'}\cos(\Vec{q}_{j}\cdot\delta\Vec{r})
\end{align*}
Thus, the autocorrelation in Eq. \eqref{eq:trig} becomes
\begin{align}
    C(\delta\Vec{r})
    =
    \frac{A ^2}{N}\sum_{j=1}^N\cos(\Vec{q}_j\cdot\delta\Vec{r})/2
    \label{eq:AC2}
\end{align}
From the autocorrelation, we can obtain the root-mean-square of the potential
\begin{align*}
    V_{\textrm{rms}}=\sqrt{\ev{(V(\Vec{r}))^2}}=\sqrt{C(0)}=A/\sqrt{2}.
\end{align*}
The fluctuation of the potential $V_{\textrm{rms}}$ is independent of the number of modes $N$ as we introduced the normalization factor $1/\sqrt{N}$ in the first place in Eq. \eqref{eq:randompot}. So far we did not assume anything about the wave vector $\Vec{q}_j$.

\begin{figure}
    \centering
    \includegraphics[width=2.5in]{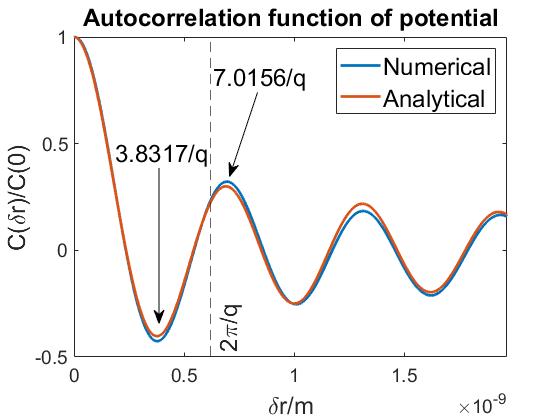}
    \includegraphics[width=3in]{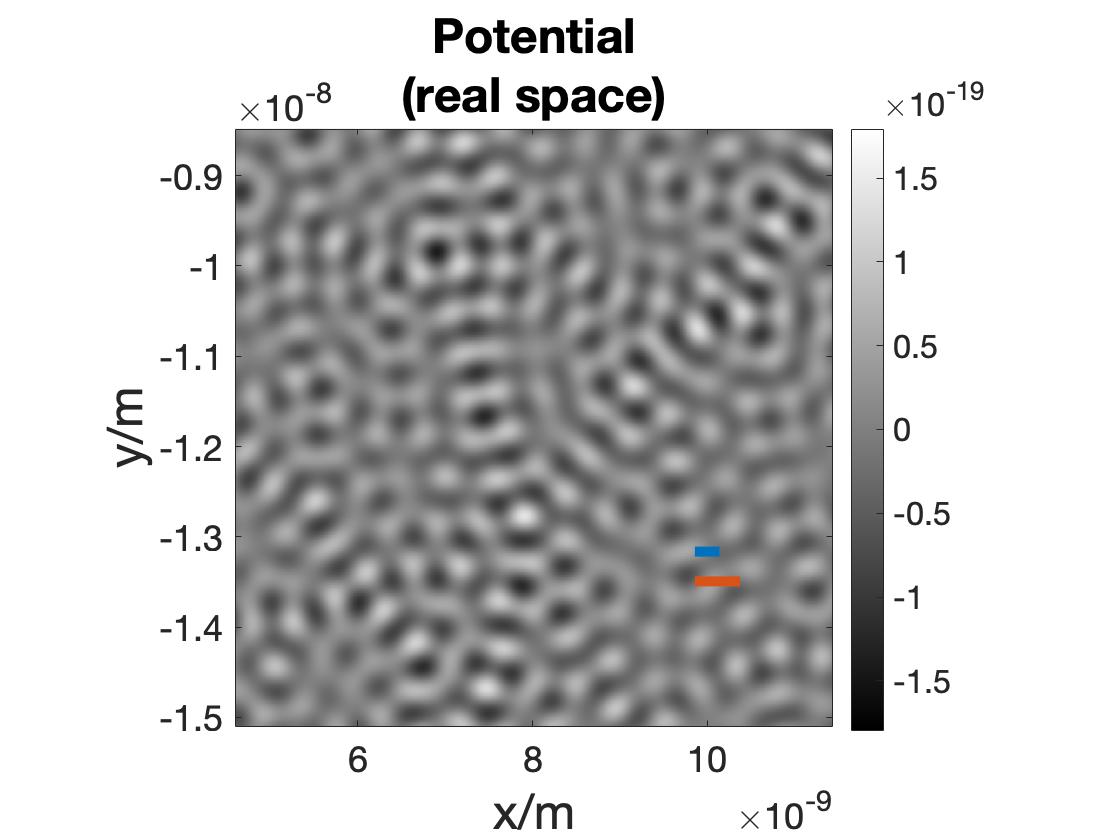}
    \caption{(left panel) The (normalized) spatial autocorrelation function of the Berry potential. Numerical result and the evaluation of the analytical expression $J_0(q\delta r)$ match well.
    Short-range order clearly exists as indicated by the pronounced peaks at $\delta r=3.8317/q$ and $7.0156/q$.
    %Two peaks at $\delta r=3.8317/q,7.0156/q$ give characteristic length scales in the potential plot in real space shown in FIG. \ref{BerryPot}.
    Note the first positive peak at $\delta r=7.0156/q$ is slightly off from the wavelength $2\pi/q$ of each sinusoid of the potential, shown as a vertical dashed line.
    (right panel) Plot of the Berry potential in real space. Thick blue and red lines show the characteristic lengths from the two peaks at $\delta r=3.8317/q$ and $7.0156/q$, respectively, of the autocorrelation function shown in the left panel. The distance between a white bump and an adjacent black dip indeed matches with the length of the blue line. Also, the distance between adjacent white bumps (or black dips) agrees with the length of the red line, showing short-range order.}
    \label{BerryACPot}
\end{figure}

Now consider the Berry potential where the moduli of $\Vec{q}_j$ are the same as $q$. In the limit of many modes $N\to\infty$ and for $\Delta\theta=2\pi/N$, the autocorrelation in Eq. \eqref{eq:AC2} becomes
\begin{align*}
    C(\delta\Vec{r})
    =
    \frac{A ^2}{N\Delta\theta}\sum_{j=1}^N\Delta\theta\cos(\Vec{q}_j\cdot\delta\Vec{r})/2
    =
    \frac{A ^2}{2\pi}\int_0^{2\pi}\dd\theta\cos(q\delta r\cos\theta)/2
    =
    \frac{A ^2}{2} J_0(q\delta r)
\end{align*}
where $J_0$ is the zeroth-order Bessel function of the first kind.
It shows not only short-range order, but also long-range order as the asymptotic expression for large distance limit
\begin{align*}
    C(\delta\Vec{r})
    \sim
    \frac{A ^2}{2}\sqrt{\frac{2}{\pi q\delta r}}\cos(q\delta r-\frac{\pi}{4})
    \text{ as }
    q\delta r\to\infty
\end{align*}
gives power-law decay $C(\delta r)\sim (\delta r)^{-1/2}$.
The Berry potential and its autocorrelation function are shown in FIG. \ref{BerryACPot}.
%Note the peaks give characteristic length scales shown in the potential plot in FIG. \ref{BerryPot}.

% There are important statistical properties of the deformation potential.
% We treat either position $\mathbf{r}$, time $t$, or phase $\varphi_{\mathbf{q}l}$ appearing in Eq. \eqref{eq:VDcl} as uniformly distributed random variables.
% Then, the deformation potential value $V_D$
% is a random variable normally distributed with mean $\mu_{V_D}$ and standard deviation $\sigma_{V_D}$ by the central limit theorem, i.e.,
% \begin{align*}
%     V_D\sim\mathcal{N}(\mu_{V_D},\sigma_{V_D}^2).
% \end{align*}
% The mean is zero $\mu_{V_D}=\ev{V_D}=0$ and the standard deviation is root-mean-square of the potential values $\sigma_{V_D}=\sqrt{\ev{V_D^2}}=V_{\mathrm{rms}}$.
% Note the average $\ev{\cdot}$ can be taken over the chosen random variables; they all give the identical results as they appear in the argument of the same cosine in Eq. \eqref{eq:VDcl}.

\section{More general potentials}
\begin{figure*}
    \centering
    \includegraphics[width=0.4\textwidth]{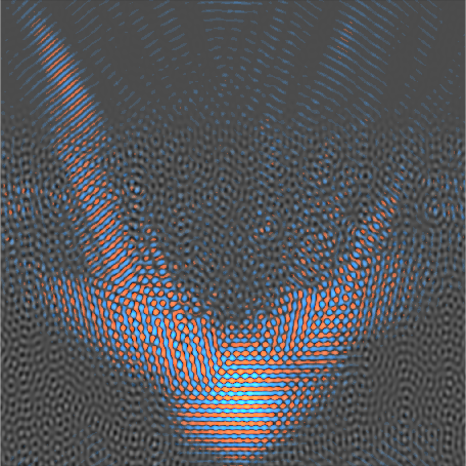}
    \includegraphics[width=0.4\textwidth]{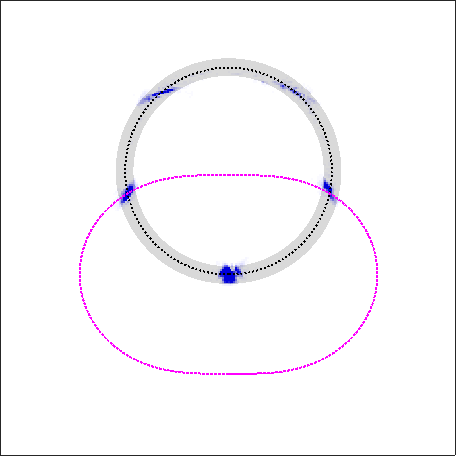}
    \caption{Real (left) and reciprocal (right) space pictures of the wave. The left panel shows the real part of the wave function, where the colormap range is chosen to show the ``Bragg wings'' better. The right panel shows the probability density of the wave with blue scale. Black dotted line shows the countour of an equal energy. The magenta dotted line shows the nonzero Fourier components of the potential where the average initial wave vector was taken as an origin. The scattered states appear at the intersections of the equal energy contour and the Fourier component of the potential. Even though the potential has a broad range of wave numbers, still a sharp scattering angle is observed since the potential forms a one-dimensional manifold in reciprocal space.}
    \label{crazypotential}
\end{figure*}

In the main manuscript, we considered the following form of a random potential
\begin{align}
    V(\Vec{r})=\frac{A}{\sqrt{N}}\sum_{j=1}^N\cos(\Vec{q}_j\cdot\Vec{r}+\phi_j),
    \label{eq:randompot2}
\end{align}
but only restricted the attention to a special case called the ``Berry potential'' where all the wavelengths of the modes are the same $q(\theta_j)=q$. However, the explanations given in the main manuscript are also valid for other random potentials rather than the Berry potential. 
Consider a potential constructed by different (angle-dependent) wave number $q(\theta_j)=q(1+0.2\cos(2\theta_j))$.
Even though the potential is constructed from a broad range of wave numbers and there is no clear length scale in it, still a sharp scattering angle is observed as shown in FIG. \ref{crazypotential}. This is because the nonzero Fourier components of the random potential lie on a one-dimensional manifold in reciprocal space (or, more formally, the random potential has one-dimensional support in reciprocal space). 
In the reciprocal space, the scattered wave indeed appears at the intersections of equal energy contour (black dotted circle) and nonzero Fourier component of the potential (magenta dotted curve). 

\begin{figure*}
    \centering
    \includegraphics[width=0.4\textwidth]{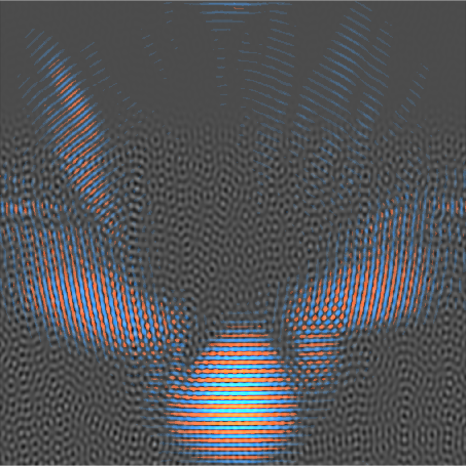}
    \includegraphics[width=0.4\textwidth]{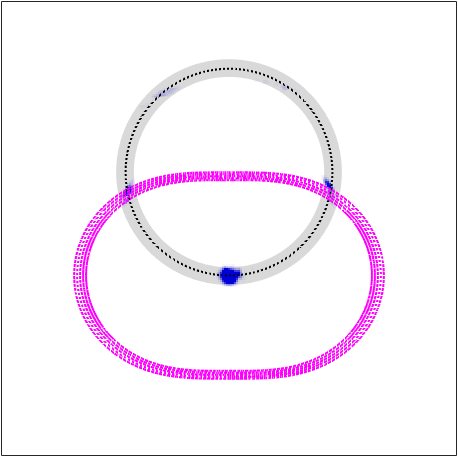}
    \caption{Real (left) and reciprocal (right) space pictures of the wave. The left panel shows the real part of the wave function, where the colormap range is chosen to show the ``Bragg wings'' better. The potential forms a ``band'' in reciprocal space and has no clear length scale in it. The right panel shows the probability density of the wave with blue scale. Black dotted line shows the countour of an equal energy. The magenta dotted line shows the nonzero Fourier components of the potential where the average initial wave vector was taken as an origin. The scattered states appear at the intersections of the equal energy contour and the Fourier component of the potential. Even though the potential has a range of wave numbers, still quite sharp scattering angle is observed.}
    \label{crazypotential2}
\end{figure*}

Furthermore, the reciprocal space analysis is not restricted to the potentials that have one-dimensional manifold of nonzero Fourier components in reciprocal space. The analysis is generally applicable to more general types of the random potential. One example is a random potential having ``bands'' in the reciprocal space. Consider the superposition of the potential just mentioned in the previous paragraph for different $q$ values (with normalization factor $1/\sum_q1$ for averaging)
\begin{align}
    V(\Vec{r})=\frac{1}{\sum_{q}1}\sum_{q}\left[\frac{A}{\sqrt{N}}\sum_{j=1}^N\cos(\Vec{q}_j\cdot\Vec{r}+\phi_j)\right]
    ,\quad
    \abs{\Vec{q}_j}=q(1+0.2\cos(2\theta_j))
    \label{eq:randompot3}
\end{align}
The superposition (average) of the potentials $q=0.96q_0,0.98q_0,q_0,1.02q_0,1.04q_0$ is shown in Fig. \ref{crazypotential2}.
Even though the potential is constructed from a broad range of wave numbers and there is no clear length scale in it, still quite sharp (but broader than one-dimensional manifold case in the previous paragraph) scattering angle is observed.

\section{Example codes for generating figures}
MATLAB codes implementing the split operator method can be found in 
\url{https://gitlab.com/public-codes/bragg-scattering-from-a-random-potential}. It reproduces figures similar to the ones in the manuscript. It contains all parameter values explicitly and we also explained how to run the program.

\bibliography{apssamp}% Produces the bibliography via BibTeX.